# Early Prediction of Alzheimer's and Related Dementias: A Machine Learning Approach Utilizing Social Determinants of Health Data

**Bereket Kindo[1] [*], Arjee Restar[1], Anh Tran[1]**
[1]IT Concepts Inc., 8201 Greensboro Drive, Suite 735, McLean, VA 22102

**Abstract**

Alzheimer's disease and related dementias (AD/ADRD) represent a growing healthcare crisis affecting over 6 million Americans. While genetic factors play a crucial role, emerging research reveals that social determinants of health (SDOH) significantly influence both the risk and progression of cognitive functioning, such as cognitive scores and cognitive decline. This report examines how these social, environmental, and structural factors impact cognitive health trajectories, with a particular focus on Hispanic populations, who face disproportionate risk for AD/ADRD. Using data from the Mexican Health and Aging Study (MHAS) and its cognitive assessment sub study (Mex-Cog), we employed ensemble of regression trees models to predict 4-year and 9-year cognitive scores and cognitive decline based on SDOH. This approach identified key predictive SDOH factors to inform potential multilevel interventions to address cognitive health disparities in this population.

**Introduction**

Alzheimer's disease and related dementias (AD/ADRD) pose an escalating medical and public health challenge, currently affecting over 6 million Americans.[1] While genetic and biological factors play crucial roles in AD/ADRD development,[2] emerging evidence suggests that social determinants of health (SDOH; i.e., social, environmental, and structural factors) can also significantly influence both risk and progression of cognitive decline.[2-5] Recent epidemiological studies have demonstrated that demographic and socioeconomic status, family history, educational attainment, and lifestyle factors (e.g., social engagements, diet/nutrition, physical activity/exercise) significantly influence cognitive decline trajectories, with estimates suggesting that up to 40% of dementia cases may be preventable through modification of these social and behavioral factors.[6-8]

*Corresponding Author: bereket.kindo@kentro.us

The relationship between SDOH and cognitive health manifests through multiple pathways across the lifespan,[9] with particular significance for marginalized and underserved populations. Often used as a proxy for socioeconomic status,[10,11] educational attainment has emerged as one of the strongest protective factors within the SDOH and cognitive health literature,[3,12-14] with one study[14] reporting that each additional year of attained education is associated with a 7% reduction in dementia risk through enhanced cognitive reserve and improved health literacy. Having employment, particularly with work that requires a higher demand for cognitive stimulation and occupational complexity, demonstrates similar protective effects, with mentally demanding occupations significantly associated with better cognitive tests scores compared to those who worked in jobs with low mental demands.[15] Additionally, lifestyle factors, including regular physical activity and social engagement, show substantial protective benefits, with socially active adults demonstrating significantly lower rates of cognitive decline compared to those who are socially isolated.[7,16-19]

Research on AD/ADRD demonstrates disparities in prevalence and outcomes across racial and ethnic groups in the United States, with particularly concerning patterns among Hispanic populations. Current epidemiological data show that AD/ADRD affects Hispanic adults at 1.5 times the rate of non-Hispanic whites,[20,21] a disparity that takes on added urgency given projections that the Hispanic population aged 65 and older will nearly quadruple by 2060.[22,23] Inequities in healthcare access and other SDOH factors may help explain this disparity. For instance, studies have found that strong familial networks and intergenerational living arrangements with multiple living children are common in Hispanic cultures and show better cognitive performance and protective benefits through increased social engagement and support systems (i.e., stronger social networks and family cohesion),[24,25] however, these benefits are often offset among families facing socioeconomic hardships,[25-27] as well as systemic barriers to education[28] and healthcare access,[29] including preventive services.[25,30] This is particularly pronounced given current English-centric healthcare system in the US, where limited Spanish language proficiency among healthcare providers and lack of culturally adapted services create substantial barriers to cognitive assessment and care delivery[31] – even in cases where family members serve as crucial intermediaries and patient advocates.[26,32] These findings underscore the importance of social engagement in generally protecting against cognitive decline while considering both SDOH impacts and population-specific contextual factors in understanding cognitive health trajectories.

Recent advances in machine learning and predictive analytics have created new opportunities for incorporating complex, large data with SDOH factors into early detection models, potentially improving risk stratification and enabling more targeted interventions in marginalized and underserved populations.[33] Previous traditional research approaches to AD/ADRD prediction have relied primarily on clinical assessments and biomarkers, with limited success in capturing the complex interplay of social and environmental risk factors.[34] However, machine learning methods have demonstrated superior capability in identifying subtle patterns of cognitive decline and integrating diverse data types, with recent studies achieving prediction accuracies of up to 85% in identifying individuals at risk for AD/ADRD several years before clinical diagnosis, showcasing the potential of these methods to enhance early detection and intervention strategies.[35] Despite these advances, most existing predictive models have focused predominantly on clinical and genetic data,[36,37] leaving significant opportunities to incorporate SDOH factors that may be particularly relevant for marginalized and underserved populations.

Recognizing this potential, the National Institutes of Health (NIH) launched the algorithms and approaches phase of "PREPARE: Pioneering Research for Early Prediction of Alzheimer's and Related Dementias EUREKA Challenge."[38] This challenge, hosted by DrivenData, aims explicitly to leverage innovative computational approaches for improving early detection and prediction of cognitive decline with a focus on examining SDOH data. In response to this data challenge, the present study utilizes longitudinal data from the Mexican Health and Aging Study (MHAS)[39] to develop and validate predictive models for cognitive decline. This work aligns with the broader NIH initiative to enhance early detection methods, particularly focusing on underrepresented populations where traditional assessment methods may be less effective or culturally inappropriate. The objectives of this study are twofold: 1) to examine the predictive utility of SDOH variables on cognitive scores at two future time points: four years post-baseline (2016) and 9 years post-baseline (2021), where the baseline is defined as 2012, the most recent year in which SDOH questionnaire data was collected, and 2) to examine SDOH predictors of cognitive decline between 2016 and 2021 using SDOH data collected at in 2012 (baseline) and 2003. By incorporating a comprehensive set of SDOH and employing advanced machine learning techniques, our study aims to address critical gaps in current prediction models while providing interpretable insights into the relative importance of various SDOH factors in cognitive decline trajectories.

**Methods:**

*Data Source*

This study uses data from MHAS, a comprehensive, nationwide longitudinal investigation of Mexican adults aged 50 and older.[39] MHAS data was collected through in-person surveys conducted in 2003 and 2012, encompasses demographics, socioeconomic factors, migration history, health status, and lifestyle information. These data serve as predictors of cognitive function for this study.

Cognitive function (i.e., cognitive scores and cognitive decline) is assessed using data from the Mexican Cognitive Aging Ancillary Study (Mex-Cog),[40] which provides in-depth cognitive assessments conducted on a subset of MHAS participants. The target variable for predictive modeling is a composite score reflecting cognitive function across seven distinct domains. This composite score is derived from detailed cognitive assessments administered in person as part of the Mex-Cog study, specifically in 2016 and 2021.

The composite score represents the cumulative points earned by an individual across seven cognitive domains, encompassing orientation, immediate and delayed memory, attention, language, constructional praxis, and executive function. These domains are assessed through various tasks, such as recalling words, understanding spatial relationships, and solving simple math problems. By exclusively using SDOH features from earlier years (2003 and 2012) to predict future composite score outcomes in 2016 and 2021, the predictive modeling maintains temporal alignment. This approach facilitates the development of widely accessible, early detection tools for assessing the risk of developing AD/ADRD. These tools can be effectively deployed in underserved communities, expanding access to early risk assessment beyond traditional, resource-intensive approaches.

*Data Preprocessing*

The training data initially contained 183 features from SDOH questionnaire results in 2003 and 2012. Missing values for categorical features were imputed by a category of 'NULL'. Missing values in variables indicating difficulty with daily living activities, such as getting dressed (adl_dress_03/12), walking across a room (adl_walk_03/12), bathing (adl_bath_03/12), and others, were imputed with a value of 0. Subsequently, these variables were summed to create an overall score reflecting the number of reported difficulties in daily living activities.

*Model Training Procedures*

This project employed a combination of CatBoost[41,42] regression models and ElasticNet regression to predict cognitive scores and cognitive decline. CatBoost is an advanced gradient boosting machine learning library that effectively handles categorical features and often outperforms existing publicly available implementations of gradient boosting on various datasets.[41,42] CatBoost introduced a permutation-driven alternative to the classic algorithm, ordered boosting to combat prediction shift caused by target leakage, and an innovative method for processing categorical features by calculating target statistics for each category based on the target values of the training samples preceding that category in a specific order.

For the CatBoost model, we employed a low learning rate of 0.001 to ensure smooth and stable convergence. To balance model complexity and generalization, the maximum depth of each individual tree was fixed at 6. The Root Mean Squared Error (RMSE) was chosen as the loss function, aligning with the evaluation metric.

To ensure robust model evaluation, 5-fold cross-validation was implemented, allowing the model to be tested on diverse subsets of the data. An early stopping mechanism was incorporated to prevent overfitting by halting training when the model's performance on a validation set stopped improving.

To obtain the final prediction of the composite score, a penalized linear model with an ElasticNet penalty was employed.[43] This model combined predictions from three separate CatBoost regression models: one predicting the composite score directly, another predicting the decline in composite score from 2016 to 2021, and a third predicting the logarithm of the composite score ratio.

*Initial Prediction of Composite Score*

To identify commonalities in SDOH that influence cognitive score prediction over both 4-year and 9-year periods, it was crucial to combine individuals with 2016 composite scores and individuals with 2021 scores with an indicator variable included to distinguish between 4-year and 9-year prediction targets. This combined approach demonstrated superior predictive performance for cognitive scores compared to separate models trained on individual subsets of the data for 4-year and 9-year predictions.

*Prediction of Cognitive Decline*

To specifically identify SDOH associated with cognitive decline, we developed CatBoost models that captured both absolute and relative changes in cognitive scores. These models were trained on a subset of the data containing individuals with both 2016 and 2021 cognitive scores. Two model variants were developed: one predicting the absolute difference between 2016 and 2021 composite scores, and another predicting the proportional change using the log-transformed ratio of 2021 to 2016 cognitive scores. This combined approach enabled the model to learn about cognitive decline in both absolute and proportional terms over time.

*Final Prediction Framework*

Final predictions for both 4-year (2016) and 9-year (2021) cognitive scores were generated by integrating outputs from the Cognitive Score Prediction Model and the two Cognitive Decline Models.

An elastic net regression model was used to combine these predictions, balancing the contributions of each model while mitigating potential collinearity.
Importantly, separate elastic net models were trained for predicting 2016 and 2021 composite cognitive scores, respectively. This approach was adopted due to the distinct effects and interpretations of the combined models across the 4-year and 9-year prediction horizons.

*Feature Importance*

To improve the interpretability of the machine learning models, SHAP (SHapley Additive exPlanations) values were employed to quantify the contribution of each feature to the predictive outcomes. SHAP values, grounded in cooperative game theory, measure the marginal impact of individual features by considering all possible combinations of feature contributions.[44] This approach allows for a detailed analysis of how specific SDOH influence the prediction of cognitive scores and decline. Visualizations such as feature importance plots were generated to highlight the relative and contextual importance of key SDOH predictors across the 4-year and 9-year prediction horizons. These insights facilitate a better understanding of the model's decision-making process and inform potential interventions targeting cognitive health.

*Evaluation Metrics*

The Root Mean Squared Error (RMSE) was selected as the primary evaluation metric to assess the predictive accuracy of the models. RMSE is defined as the square root of the mean squared difference between observed and predicted values, offering an intuitive measure of model error. It is particularly suitable for this study as it emphasizes larger errors, which are critical when predicting cognitive outcomes where small deviations may indicate significant clinical implications.[45] Lower RMSE values indicate a closer alignment between the predicted and actual cognitive scores, validating the model's reliability.

**Results**

Our analysis of the Mexican Health and Aging Study (MHAS) data revealed significant associations between various social determinants of health (SDOH) and cognitive outcomes. Using CatBoost regression models combined with elastic net integration, we identified key predictors of both cognitive scores and decline trajectories over 4-year (2016) and 9-year (2021) periods.

*Model Performance and Feature Importance.* The CatBoost regression models demonstrated strong predictive performance, with education level (edu_gru_12) emerging as the most influential feature

(importance score: 29.68), followed by age (age_12; importance score: 6.87) and number of activities of daily living limitations (act_time_use_sum; importance score: 5.04). These three features accounted for a substantial portion of the model's predictive power, highlighting their crucial role in cognitive health trajectories.

*Demographic and Educational Predictors*. Age and educational attainment emerged as primary predictors of both cognitive scores and decline rates. Lower educational attainment consistently predicted both lower cognitive scores and accelerated decline. Intergenerational educational effects were evident, with mother's educational level showing a positive association with cognitive scores. These findings underscore the enduring influence of educational factors across generations.

*Health and Physical Function.* Several health-related factors demonstrated significant associations with cognitive outcomes:

- Higher number of illnesses predicted both lower cognitive scores and accelerated decline
- Regular exercise (three times per week) was protective against cognitive decline
- Higher BMI (overweight category) was associated with higher cognitive scores, contrary to some previous findings
- Greater difficulty with daily activities predicted lower cognitive scores, suggesting an important relationship between physical function and cognitive health

*Social and Family Factors.* Social and family characteristics showed notable associations with cognitive outcomes:

- Higher number of living children predicted better cognitive scores
- Having a female spouse was associated with both higher cognitive scores and lower decline rates
- Less frequent social contact (monthly versus weekly visits with friends and relatives) predicted lower cognitive scores
- Higher weight over personal decisions was associated with both lower cognitive scores and greater decline

*Environmental and Lifestyle Factors.* Housing quality and lifestyle engagement emerged as significant predictors:

- Floor material quality (specifically wood mosaic) showed a complex relationship, predicting higher initial cognitive scores but greater subsequent decline
- Engagement in cognitive activities (crosswords, puzzles, games) predicted higher cognitive scores
- Activity-related time use emerged as the third most important predictor overall (importance score: 5.04), emphasizing the significance of lifestyle factors

*Temporal Patterns.* Our analysis of both 2003 and 2012 data revealed that more recent measurements (2012) generally showed stronger predictive power than earlier ones (2003), though historical data remained relevant for understanding cognitive trajectories. The age variable from 2003 (importance score: 3.02) retained significant predictive value, suggesting cumulative effects over time.

*Model Integration.* The elastic net regression successfully integrated predictions from multiple models, including direct cognitive score prediction and two approaches to cognitive decline (absolute difference and logarithmic ratio). This integration improved the robustness of our predictions across different time horizons and population subgroups. Root mean squared errors of our approach compared to elastic-net penalized regression are in Table 1.

Table 1. Root mean squared error (RMSE) of the final "integrated" model compared to

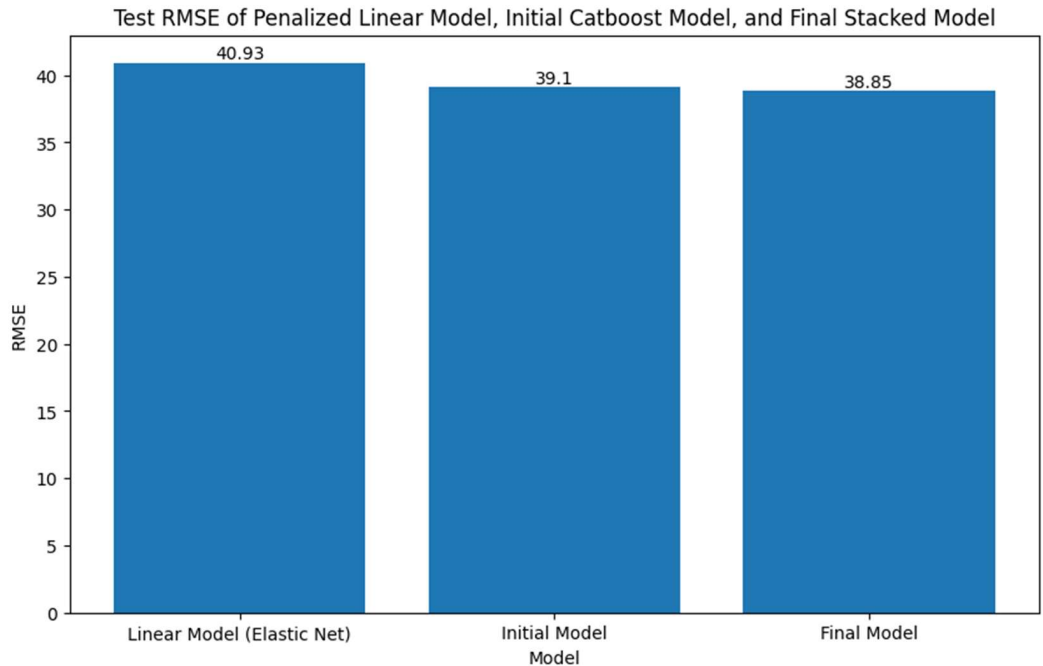

**Discussion**

    Our findings support that cognitive health trajectories among Hispanic older adults are shaped by an array of SDOH factors, encompassing demographic, family composition and history, educational attainment, and lifestyle factors (e.g., social engagements, diet/nutrition, physical activity/exercise). More importantly, the predictive patterns that emerged from our analysis demonstrate both protective and risk factors for cognitive scores and decline. These results highlight the multifaceted nature of cognitive health determinants and underscore the importance of considering individual, social, and structural factors in understanding cognitive decline risks, particularly in marginalized populations. Taken together, the patterns we identified suggest that interventions targeting multiple SDOH domains simultaneously may be more effective than singular approaches, echoing previous calls for multi-prong interventions and the need for more nuanced approaches to risk assessment and intervention design.

    Looking closely at each specific predictor, our analysis identified age and educational attainment as primary predictors of both cognitive scores and decline rates, aligning with longitudinal studies showing that additional years of education significantly reduce dementia risk[13] and that cognitive decline accelerates in later years.[46] Specifically, age-related cognitive decline could highlight how cumulative

exposure to social and environmental stressors over time, in addition to biological reasonings, may compound age-related vulnerabilities in underserved populations.[46] Educational attainment, meanwhile, as supported by previous literature,[3,12-14] emerges as a powerful marker of socioeconomic protective effect, likely as a proxy to broader lifetime access to health-promoting resources, enhanced social and economic opportunities, and greater ability to navigate healthcare systems.[11] This finding in our models underscores how early-life educational disparities, often stemming from systemic inequities in educational access, can shape cognitive health trajectories throughout the life course.[10,11] This pattern is particularly concerning for Hispanic populations who have historically faced barriers to educational opportunities due to linguistic, cultural, and structural factors,[47,48] suggesting that addressing educational equity represents a crucial upstream intervention point for reducing cognitive health disparities in this community.[49,50]

Lifestyle and physical health-related factors in our study – in particular, BMI, exercise, engagement in daily activities, and number of illnesses – showed significant predictive relationships that both align with and challenge existing research. For instance, our observation that overweight BMI predicts higher cognitive scores contrasts with meta-analyses showing increased dementia risk with higher BMI,[51] potentially reflecting the "obesity paradox" previously documented in Hispanic populations.[52,53] This paradox, where overweight status appears protective rather than harmful, may be explained by both the limitations of BMI thresholds derived from predominantly white populations that do not account for racial/ethnic differences in body composition and the complex interplay of immigration status, acculturation level, and dietary patterns that moderate BMI-health relationships in Hispanic populations.[54,55] Additionally, the protective effect we observed might also reflect socioeconomic factors, as higher BMI in some Hispanic communities may correlate with better food security and access to resources,[56,57] which could indirectly benefit cognitive health through improved overall nutrition and reduced stress.[58] Our findings also align with previous intervention studies[59] that show the protective effect of regular exercise on cognitive decline and suggest the importance of maintaining physical activity throughout aging. Similarly, we also found that multiple illnesses predict lower cognitive scores, which aligns with longitudinal studies showing that multimorbidity accelerates cognitive decline.[60] The association between greater difficulty in activities of daily living (ADLs; e.g., getting dressed, eating, walking from one side of the room to another, getting in and out of bed, using the toilet) and lower cognitive scores likely reflects a bidirectional relationship where physical limitations may reduce opportunities for physical and social engagement, while cognitive decline itself may impact functional abilities and independence.[61,62] This finding, coupled with our observation that regular exercise predicts lower cognitive decline, underscores the interconnected nature of physical and cognitive function in aging, suggesting that early interventions to maintain physical function and promote regular exercise may help preserve both physical independence and cognitive health. Lastly, the relationship between illness burden and cognitive outcomes appears to be particularly strong in our Hispanic cohort, suggesting potential disparities in healthcare access, successful treatment management, and overcoming physical and chronic health conditions. Taken together, these complex relationships between lifestyle and physical health-related factors and cognitive outcomes highlight the need for culturally-informed approaches to health promotion and AD/ADRD prevention.

Our findings on social engagement and family structure reveal nuanced relationships that extend current understanding. The protective effect of having more living children on cognitive scores supports systematic reviews showing that more extensive social networks generally reduce dementia risk.[16] Our observation that monthly (versus weekly) social contact predicts lower cognitive scores further demonstrates the importance of social engagement, particularly the frequency of social interaction for maintaining cognitive function. The novel finding that having a woman spouse predicts better cognitive outcomes adds to emerging research on gender differences in caregiving quality,[63] though, it warrants careful interpretation through multiple lenses. This association may reflect several underlying mechanisms. First, gender differences in life expectancy likely play a crucial role, where men in our sample may be more likely to have living spouses compared to women,[64] potentially creating a selection effect that influences our observed associations (i.e., that there are more respondents with a living woman spouses in the sample compared to respondents with a living man spouse). Second, well-documented gender disparities in caregiving quality and approach may be particularly relevant, with evidence suggesting women typically provide more comprehensive and preventive care, including greater attention to cognitive stimulation, medication adherence, and healthcare appointment maintenance.[63,65] Women caregivers also tend to report higher levels of empathy and emotional support provision, which are factors that may buffer against cognitive decline.[66-68] Third, this finding may reflect broader social isolation patterns, where having a living spouse, regardless of gender, may protect against loneliness and cognitive decline through increased daily social interaction, shared activities, and mutual support.[69-71] In Hispanic communities specifically, these spousal effects may be amplified by cultural values that emphasize strong marital bonds and family-centered care practices.[29] Future research should further investigate the impact of spouses' gender on cognitive functioning. Altogether, these findings suggest the importance of considering cultural context in understanding social support mechanisms to prevent AD/ADRD. Future research should look into delineating the varying strength of social engagement effects across different cognitive domains, given that our findings indicate that social interaction may differentially affect various aspects of cognitive function.

There were also some SDOH-related socioeconomic indicators demonstrating complex temporal relationships that inform our understanding of cognitive aging worth noting. Our observation that higher-quality housing (e.g., floor wood quality) predicts better initial cognitive scores but greater cognitive decline rates parallels research showing that socioeconomic advantages may delay but not prevent cognitive decline.[72] This finding suggests that the physical environment may operate through multiple pathways to influence cognitive health, though future research is needed to better understand how specific housing characteristics interact with other SDOH factors to affect cognitive trajectories, particularly in communities facing housing instability or environmental hazards. Additionally, longitudinal studies examining how changes in housing quality over time affect cognitive outcomes could help clarify these complex relationships.

Taken together, these findings collectively underscore the importance of considering multiple SDOH domains in understanding cognitive health trajectories among Hispanic older adults. The complex associations between age, educational attainment, lifestyle and physical health-related factors, social engagement, and physical environment factors align with emerging frameworks emphasizing the multifaceted nature of cognitive decline risk. Our results particularly support calls for culturally adapted interventions that leverage existing family and social networks while addressing social and systemic barriers to healthcare access. Large-scale intervention studies have demonstrated that multicomponent approaches addressing multiple risk factors can significantly reduce cognitive decline,[73] and could be a promising approach to Hispanic populations. These patterns suggest the need for comprehensive intervention strategies that address multiple SDOH domains simultaneously.

*Limitations*

Our study has several important limitations that should be considered when interpreting the findings. While our sample provides valuable insights into Hispanic aging populations, it may not fully represent the diversity of all populations at risk for AD/ADRD in the US,[74] potentially limiting generalizability. The temporal gaps between predictor measurements (2003/2012) and outcomes (2016/2021) may not capture all relevant changes in SDOH factors, particularly given the dynamic nature of social and environmental factors.[75] Additionally, while our model demonstrates strong predictive performance, the complex interactions between SDOH and cognitive decline may not be fully captured by our current analytical framework, suggesting the need for future research employing more sophisticated approaches to modeling these multifaceted relationships. Finally, our reliance on self-reported measures for several predictors may introduce recall bias, a limitation that is particularly relevant given the study's focus on cognitive health outcomes. Future studies should consider incorporating objective measures and more frequent assessment points to address these limitations.

**Conclusion**

Our successful application of machine learning approaches to predict cognitive outcomes using SDOH data aligns with NIH's strategic priorities for advancing early detection of AD/ADRD in diverse populations. Our study identified key SDOH predictive factors, demonstrating that machine learning techniques are applicable to examine SDOH data and can effectively predict cognitive outcomes beyond traditional clinical and genetic markers.[33] The ability of our models to identify complex patterns, such as how education and regular exercise protect against cognitive decline while ADL difficulties and limited social engagement predict poorer outcomes, supports the NIH's emphasis on innovative computational approaches for improving early detection. The successful integration of multiple SDOH domains in our predictive models suggests potential for broader application in clinical settings, particularly for Hispanic populations where socioeconomic factors and cultural contexts play crucial roles in cognitive health trajectories. Our findings also highlight the importance of maintaining model interpretability while capturing complex social determinants, as understanding these relationships is crucial for developing targeted interventions. Furthermore, the strong predictive performance of SDOH factors, especially those that are potentially modifiable like physical activity and social engagement, suggests that future clinical risk assessment should incorporate these social and environmental measures to better identify and address cognitive health disparities.

# Appendices

## Appendix A: Individual Level Prediction Explanation Examples

Figure A.1: Prediction Explanation validation *uid: bpja*

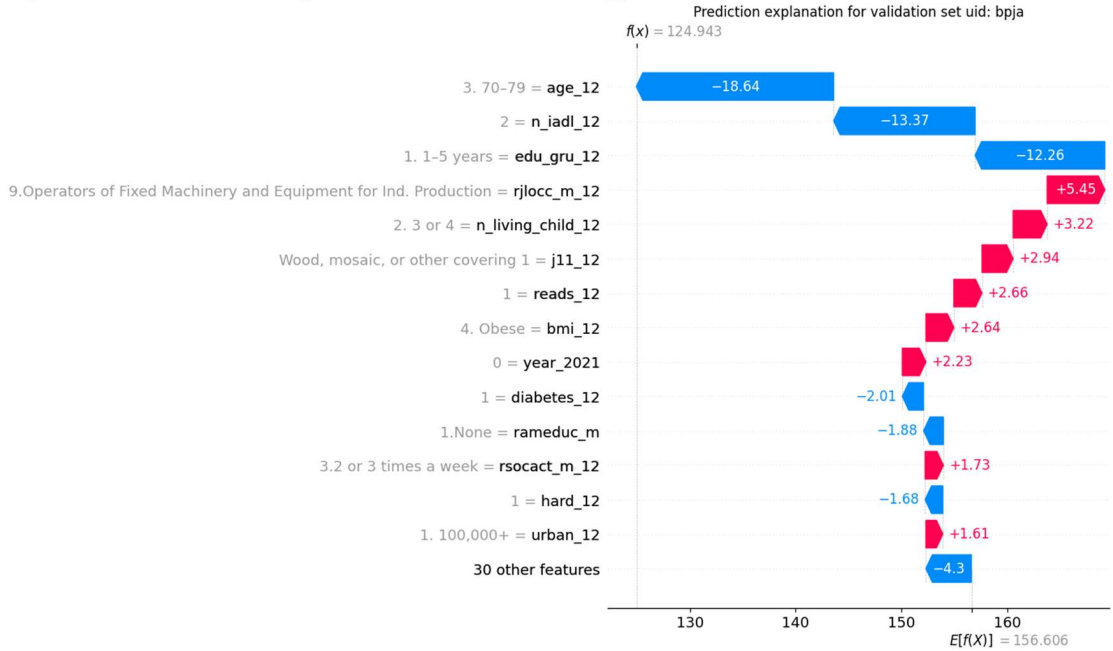

Figure A.2: Prediction Explanation validation *uid: bqif*

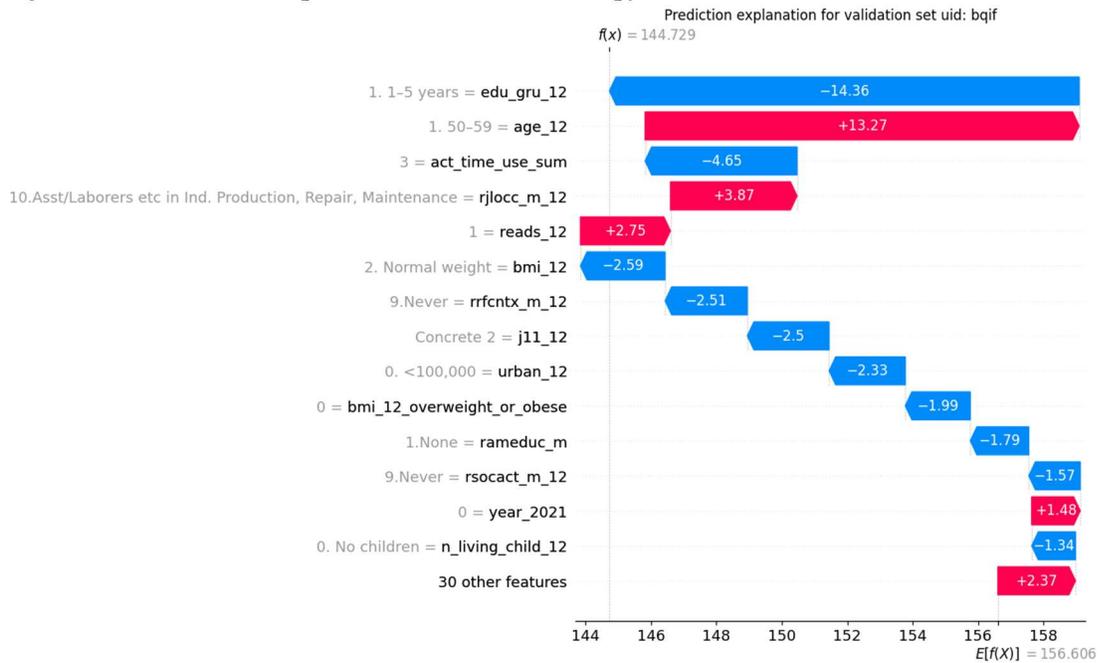

Figure A.3: Prediction Explanation validation *uid: bqxj*

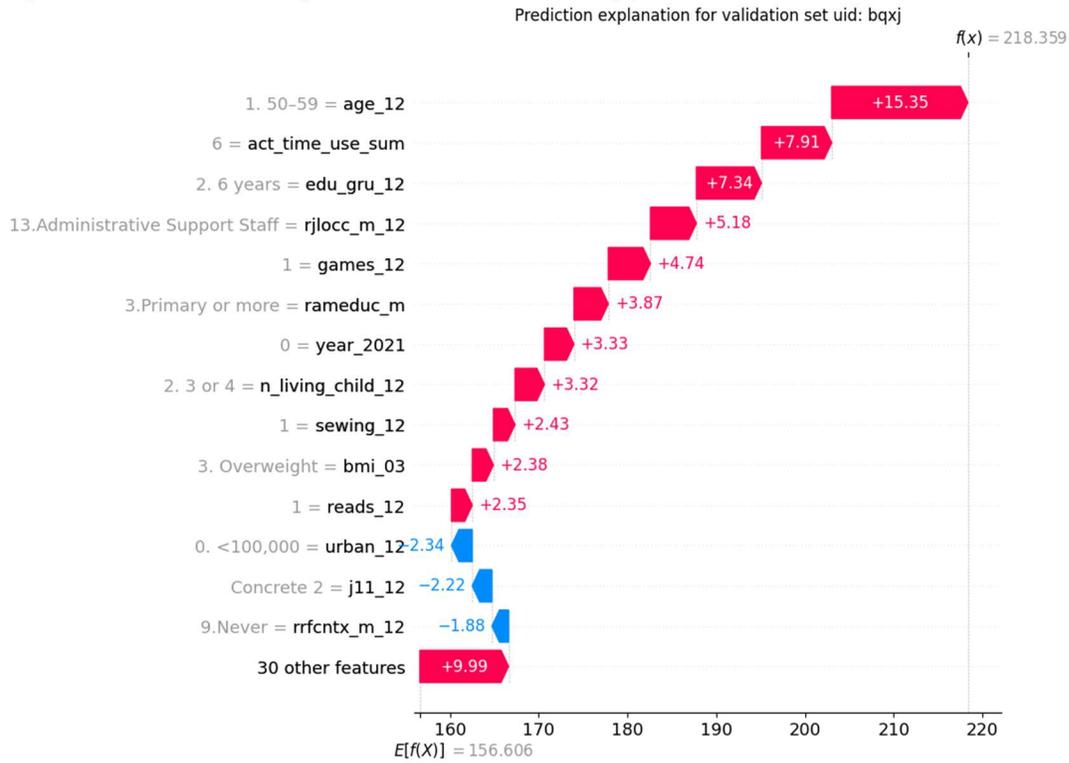

Figure A.4: Prediction Explanation validation *uid: btdk*

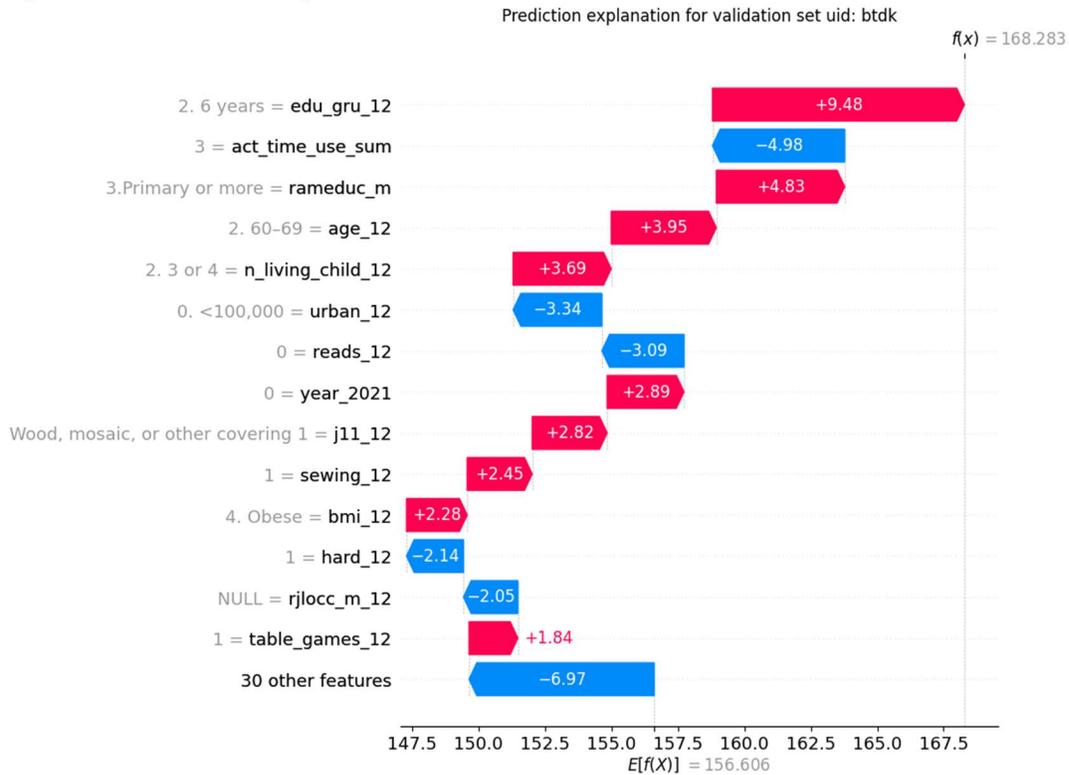

Figure A.5: Prediction Explanation validation *uid: bupw*

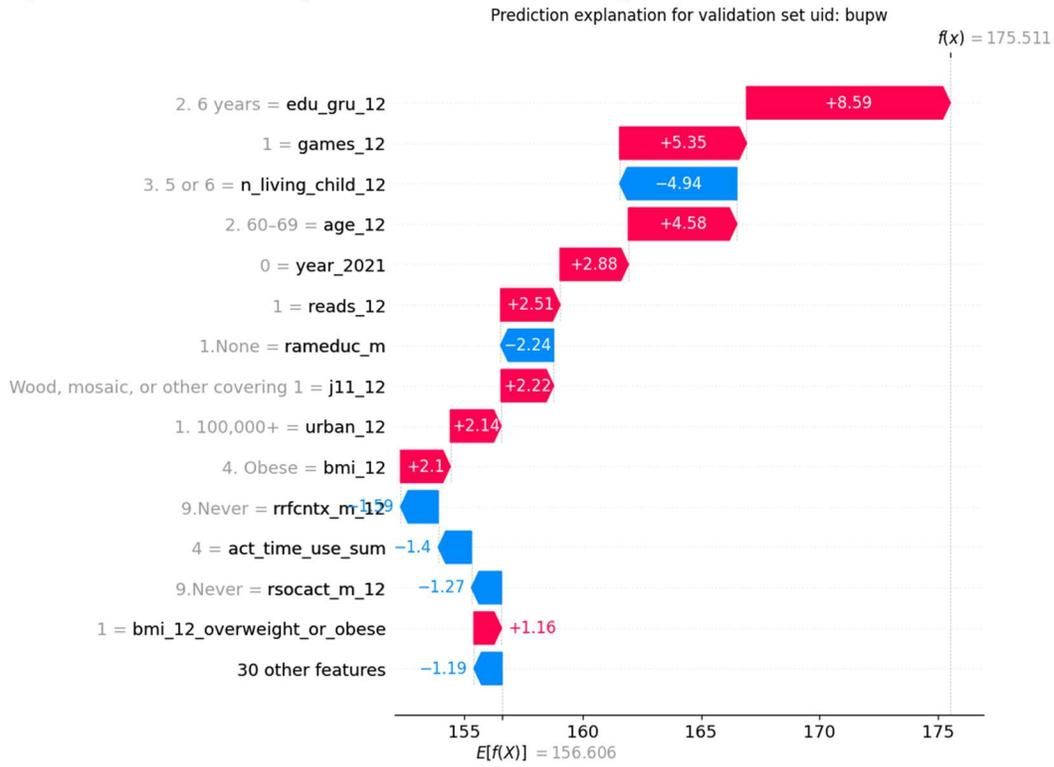

**Appendix B: Main Effect SHAP Value Estimates of Key Predictors of Composite Score**

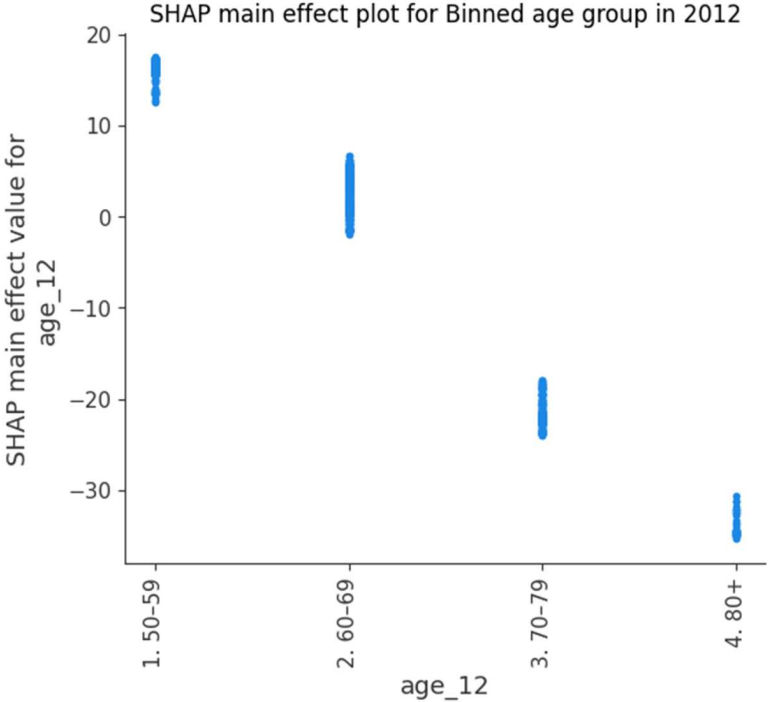

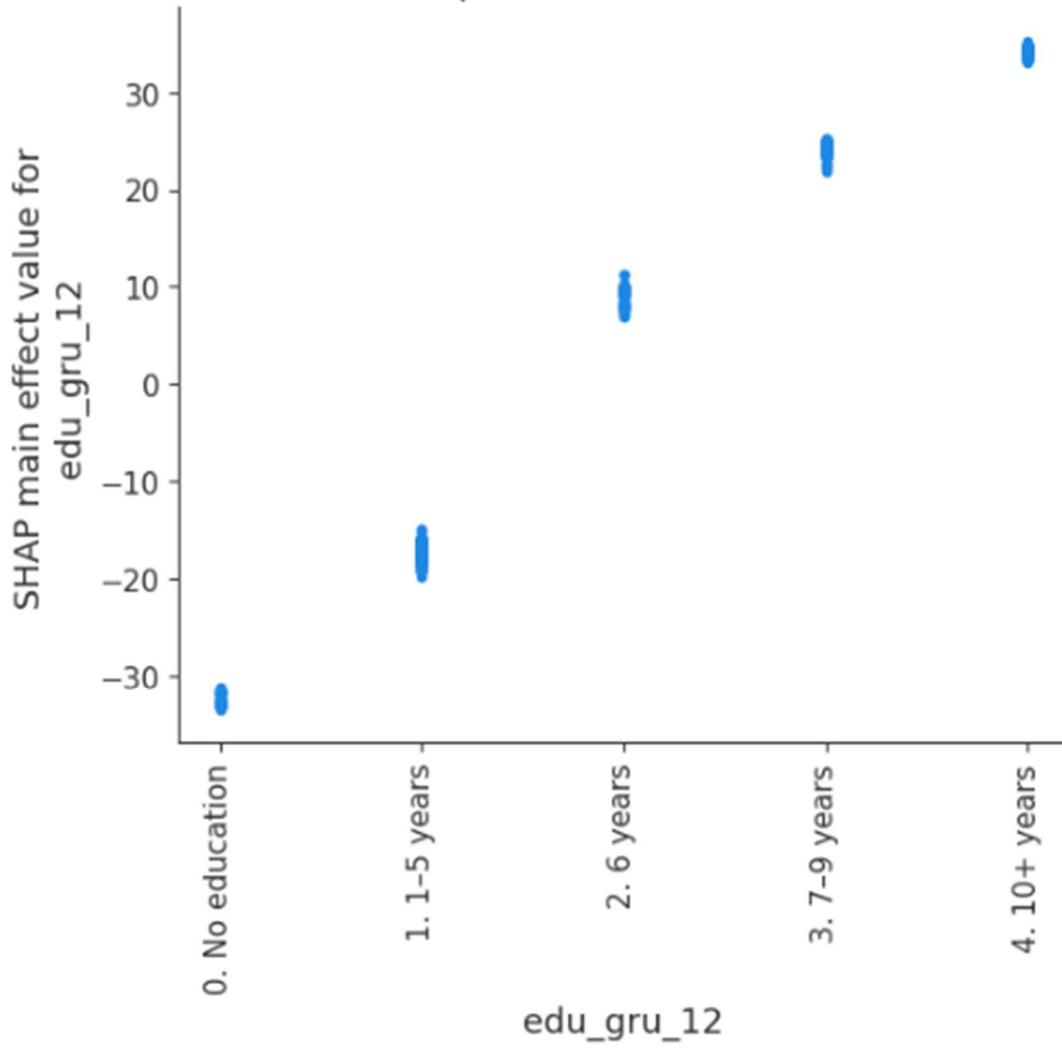

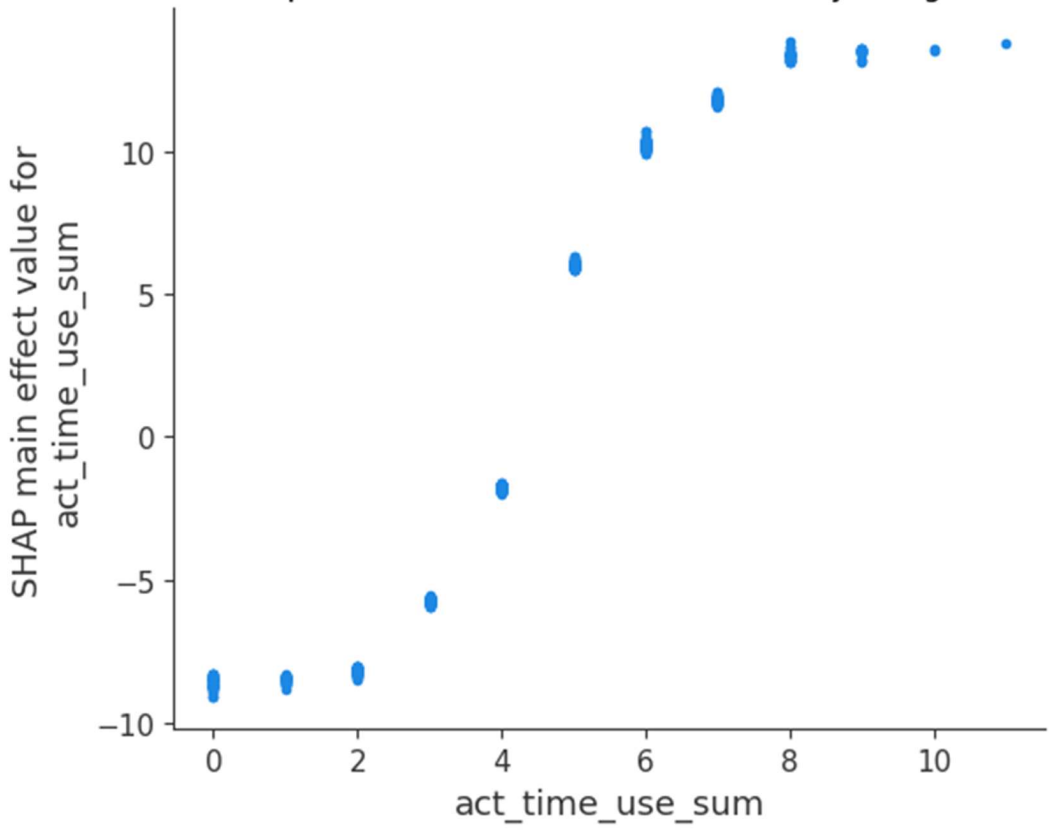

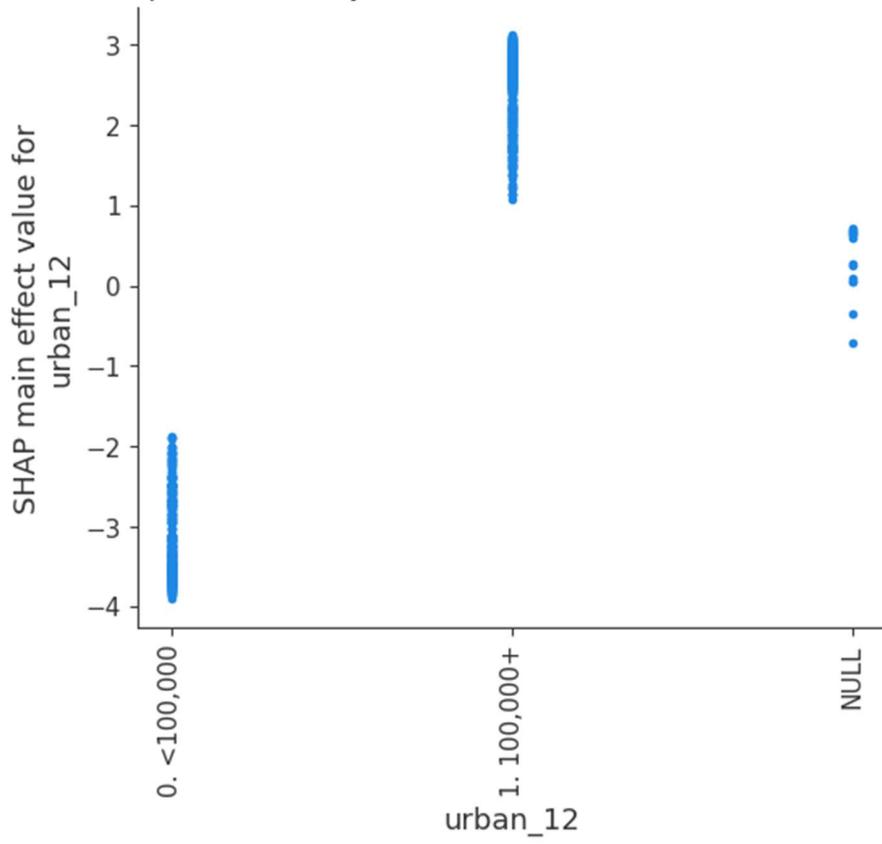

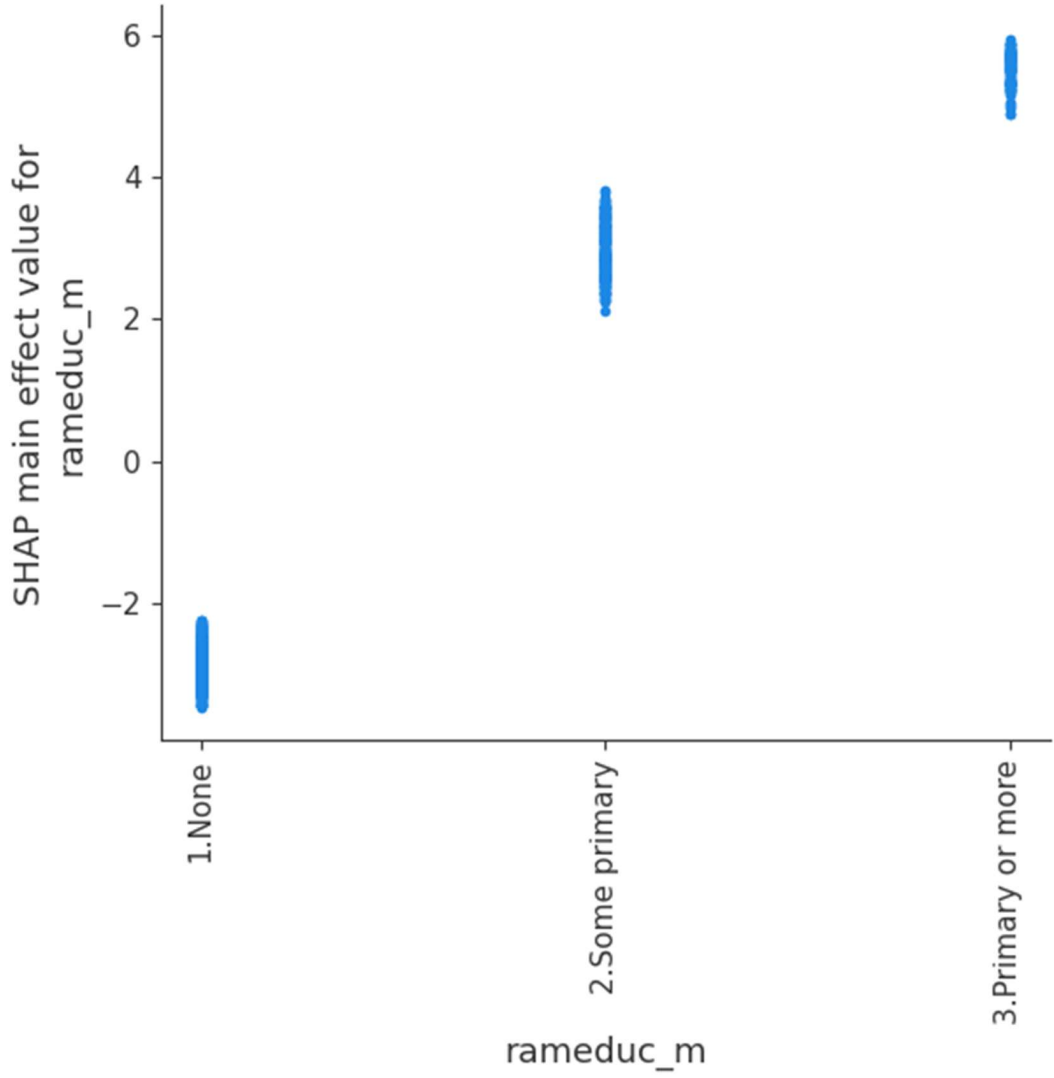

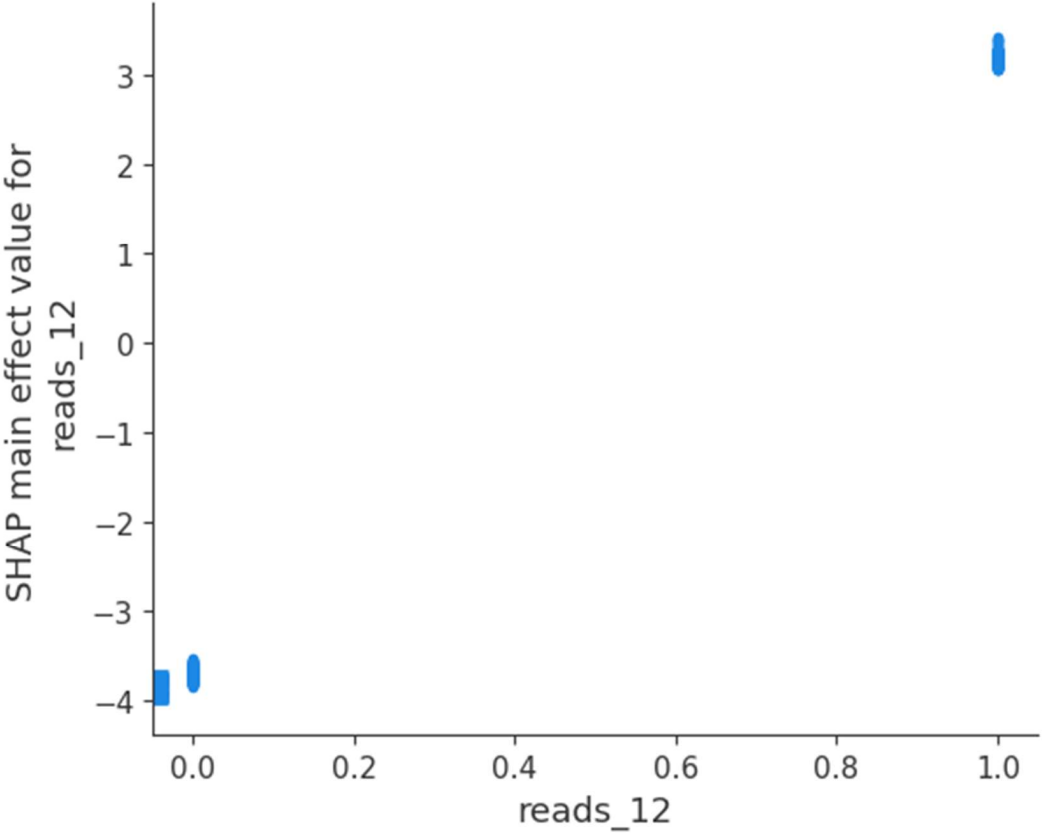

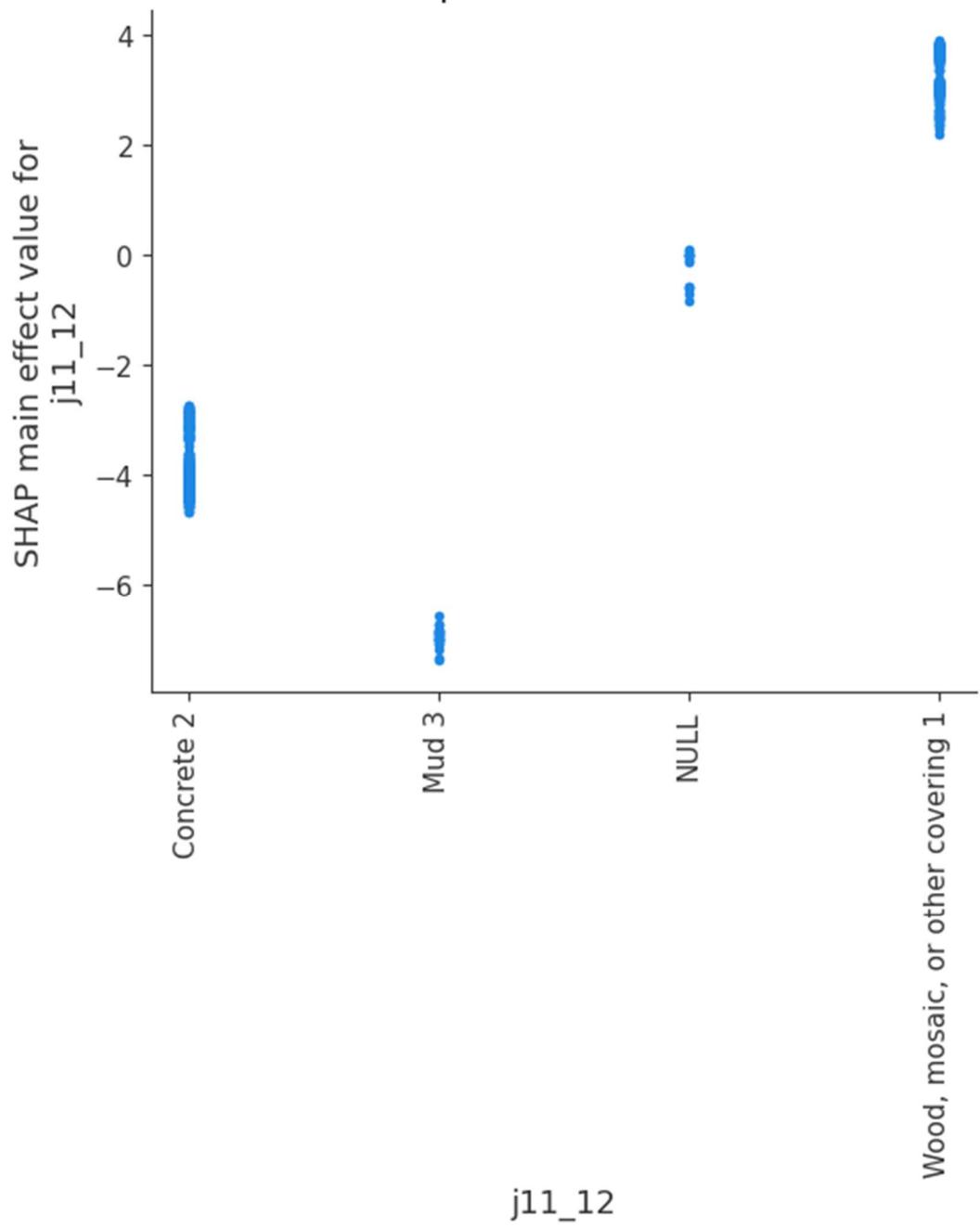

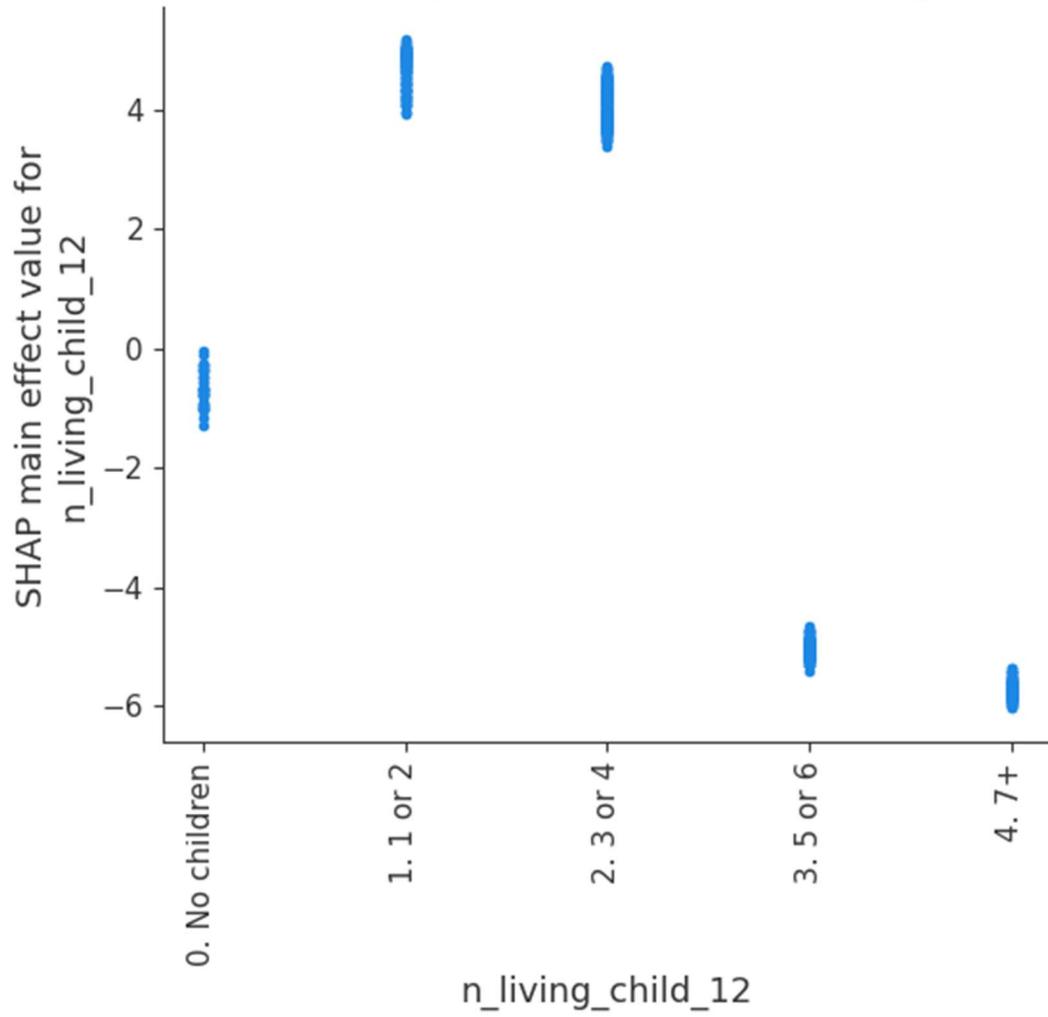

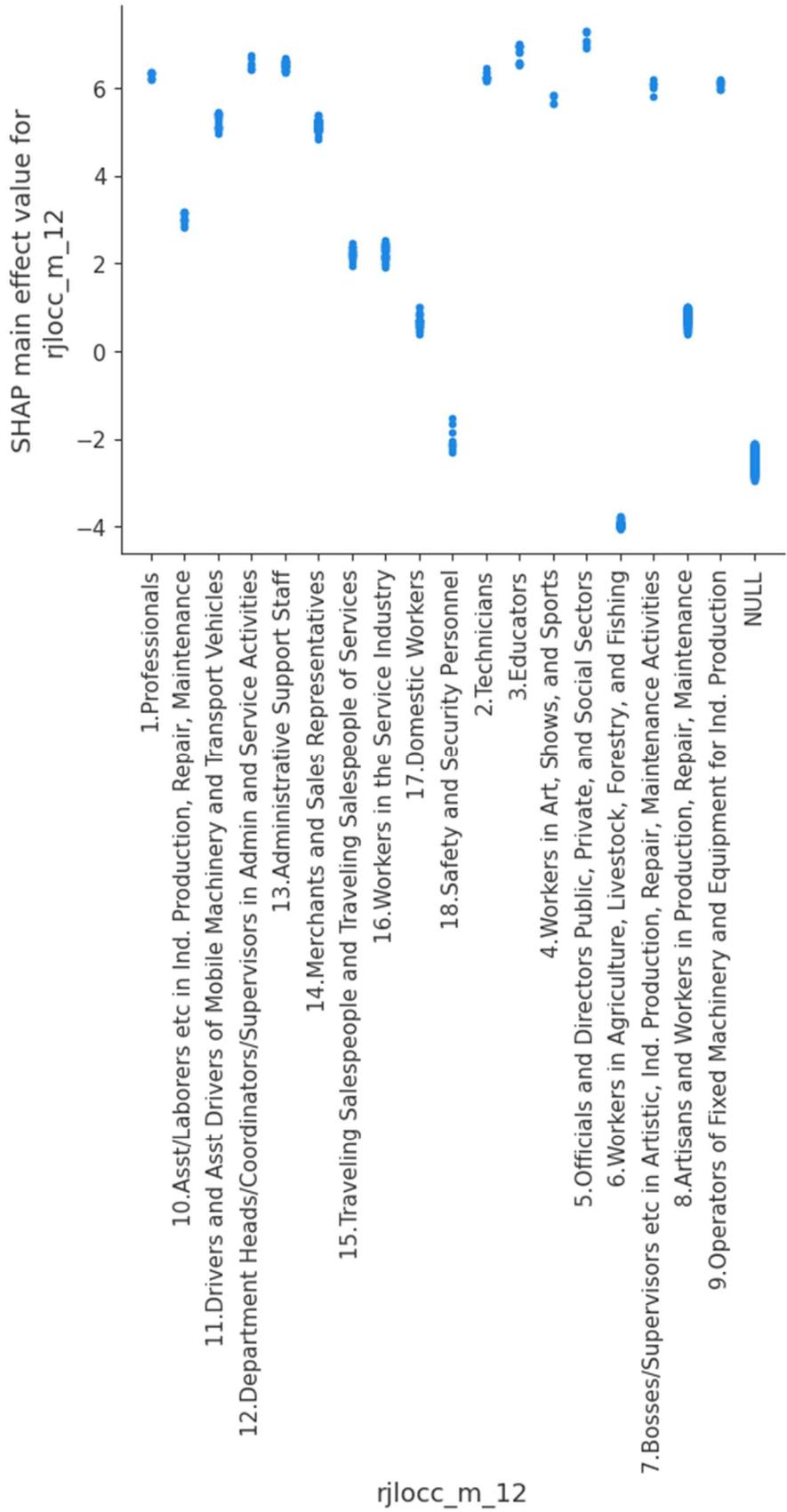